\documentclass[10pt,preprint]{aastex}
\usepackage{color}
\begin{document}

\title{Ultrahigh Energy Cosmic Rays: A Galactic Origin?}
\author{David Eichler$^{1}$, Noemie Globus$^{2,1}$, Rahul Kumar$^{3,1}$ and Eyal Gavish$^{1}$}
\altaffiltext{1}{Department of Physics, Ben-Gurion University, Be'er-Sheba 84105, Israel}
\altaffiltext{2}{Racah Institute of Physics, The Hebrew University, 91904  Jerusalem, Israel}
\altaffiltext{3}{Department of Astrophysical Sciences, Princeton University, Princeton, New Jersey 08544, USA}

\begin{abstract}
It is suggested that essentially all of the UHECRs we detect, including those at the highest energy, originate in our Galaxy.  It is shown that even if the density of sources decreases with Galactic radius, then the anisotropy and composition can be understood. Inward anisotropy, as recently reported by the Auger collaboration can be understood as drift along the current sheet of UHECRs originating outside the solar circle,  as predicted in Kumar and Eichler (2014), while those originating within the solar circle exit the Galaxy at high latitudes.
\end{abstract}

\section{Introduction}
Could ultrahigh energy cosmic rays, even at the highest energies, be Galactic in origin? The widely held belief to the contrary is based on the argument that, if they are of Galactic origin, they could hardly be as isotropic as presently observed.  However, we suggest that the isotropy can be understood in the context of Galactic production if a sufficiently careful treatment of cosmic ray propagation is undertaken. In particular, we cite an earlier work (Kumar \& Eichler, 2014) in which both anisotropic diffusion and drift are taken into account.  

In this picture, the UHECR energy spectrum displays a cutoff at $\sim 5 \cdot 10^{19}$ eV not because of any energy threshold for a high energy process (such as a photopion production threshold or photodissociation threshold), but simply because the sources cannot accelerate nuclei much beyond that value. This  so called "maximal energy" scenario has been proposed before in the context of extragalactic UHECR (for a review, see Olinto 2012) but it has been noted [e.g.  by Aloisio, Berezinsky and Blasi (2014)] that the presence of intermediate elements such as C,N,O, implies an "anti-ankle" rather than an ankle for spectral indicies that are about as steep, or steeper, than $-2$, as is expected of shock acceleration. On the other hand, the spectrum of {\it escaping} UHECRs may be sufficiently hard that, if the non-escaping particles are discounted because they suffer adiabatic losses, the maximal energy scenario may be possible. 

The actual Galactic source of the UHECR is a separate question. However, the analysis depends on whether their sources provide a steady input (i.e. frequent, low yield events) or infrequent, high yield bursts that occur less than once per UHECR escape time from the Galaxy. The latter can produce spectra that vary in time, leaving an extra free parameter. In this paper we consider time average spectra, this is done in ignorance of conceivable time variability, and motivated by the  desire to at least make a plausible case. But it should be kept in mind that a wider range of possible spectra and compositions is also possible, and further investigations are planned.

If the Galactic sources of UHECR are compact then UHECR emerging from a compact Galactic source may undergo photodissociation at the source.  In this case, it is also  necessary to consider 
concrete models of the compact object to quantitatively estimate the composition of the emergent UHECR. 
Shocks associated with baryonic outflows of gamma-ray bursts (GRB)  are  a  candidate for the acceleration of UHECR 
(Levinson and Eichler 1993; Milgrom and Usov 1995, Vietri, 1995; Waxman 1995). Levinson and Eichler specifically proposed Galactic GRBs as sources of UHECR, realizing that the total power in extragalactic GRBs, if distributed smoothly,  would probably be too small to account for an extragalactic flux at the observed levels. We shall proceed to adapt Galactic GRBs as a concrete model for UHECR sources.

\section{Model}

In our modeling of the Galactic UHECR spectrum, we use detailed calculations of UHECRs acceleration at GRB internal shocks by Globus et al. (2015). 
Two important aspects are taken into account during the calculations  of the accelerated particle spectrum: 1)  the escape of the particles from the acceleration site, which behaves as a high pass filter since only particles in the weak scattering regime manage to escape from the magnetized region upstream of the shock. Thus, particles escaping from the GRB environment  have a hard spectrum (approximately  $E^{-1}$); and 2)  the energy losses and photodissociation processes, which limit the maximal energy and produce a large amount of neutrons.  The result (a key feature for the interpretation of the ankle) is that UHE protons have a much softer spectrum than other UHE nuclei. 

There is one important difference  between the compositions of shock accelerated particles within the Galaxy and of extragalactic ones: At a given energy $E$, heavy nuclei have {\it smaller} rigidity $R\simeq E/Q$ by the factor $Q$, the nuclear charge. So if the escape rate from the Galaxy goes as $R^{\alpha}$, where at low energies $0.2 \lesssim \alpha \lesssim 0.4$, then the heavy nuclei are enhanced relative to protons  by an additional factor $Q^{\alpha}$.  This factor is besides the factor $A^{2+p}$ that obtains when going from relative abundances at a given energy/nucleon to a given total nuclear energy (where the production spectrum goes as $E^{-2-p}dE$). When using the relative abundances at 10 GeV, the value of  $A^{2+p}$ should also be factored in at that energy, and this provides the relative abundance at a given total energy per nucleus up to the exponential cutoff at high energy, (which for protons is 1 - 3 EeV).  
 Test particle simulations show that the escape rate from the Galaxy at near-ankle energies is nearly proportional to $E/Q$ (Figure 5 of Kumar and Eichler, 2014), so we consider values in the range of $\alpha \sim 0.9$.

\section{Results}

Figure 1 shows the predicted UHECR total energy spectrum resulting from the Fermi acceleration of a mixed composition of cosmic-rays at GRB internal shocks, as calculated by Globus et al (2015). 
The top panel shows the emission expected for a single GRB, which is assumed to have an isotropic equivalent  prompt $\gamma$-ray output of $5 \cdot 10^{53}$ erg.      
The wind duration is 2s.    The metallicity is assumed to be 10 times larger than  the Galactic CR source composition. The contribution of different groups of nuclear species is shown. The normalization is obtained by integrating over the whole shock propagation, and assuming that the dissipated energy at the internal shocks is equally shared between the electrons, the cosmic-rays and the magnetic field (see Globus et al. 2015 for more details). 
   The evolution of the composition is a consequence of the acceleration process and can be interpreted in a natural way if the UHECR source can accelerate protons only up to an energy $E_{\max}(\mathrm{p})$ (here a few $10^{19}$~eV), and other nuclei up to the same rigidity, i.e. an energy $E_{\max}(^{A}_{Q}\mathrm{X)}= Q\times E_{\max}(\mathrm{p})$ (Allard et al. 2008).

The lower panel shows the cosmic ray flux expected at the Earth assuming that the source is the Galactic GRB of the top panel. 
The normalization is chosen to fit the data, but let us work backwards, following Eichler and Pohl (2011), to see the requirements it makes on the Galactic sources of UHECRs:   If the sources were steady, distributed  homogeneously in the Galactic equatorial plane, and had a power per unit area of $\Sigma$, and if the UHECR emanate isotropically from them,  then the net  all-sky flux at Earth $F_{net}$ would be given by 

\begin{equation}
F_{net}=\Sigma\int_{D_{min}}^{D_{max}} \frac{2\pi DdD}{4\pi D^2} = \frac{1}{2}\Sigma \ln \left(\frac{D_{max}}{D_{min}}\right),
\end{equation}
where $D_{min}$ should be chosen to be the distance to the closest source, and may depend on assumptions about the discreteness of the sources.  (In the source continuum limit, it would be the thickness of the Galactic disk, below which the planar approximation breaks down.) However, each cosmic ray can cross a sphere containing the Earth many times, so the the measured flux $F(E)$ at a given energy $E$ (formally the energy flux in UHECRs at energy $E$ per unit logarithm in $E$) is actually of the order of  

\begin{equation}
F(E)\simeq\Sigma\int_{D_{min}}^{D_{max}} \frac{2\pi DdD}{4\pi  a(D,E)D^2}, 
\label{calc}
\end{equation}
where $a(D,E)$ is the anisotropy at Earth of cosmic rays originating from a point source at distance $D$. The inexact equality in Equation \ref{calc} is due to the uncertainty due to anisotropic transport, the issue of which we resolved with numerical calculation of  UHECR trajectories in a realistic Galactic magnetic field.   In Figure 2 we display in the upper panel the net all-sky flux $F_{net}$ at $E=2.4$ EeV contributed by a Galactocentric ring of sources in the Galactic plane at a radius $R$ from the Galactic center, where the flux has been calculated numerically using simulated test particle trajectories (Kumar and Eichler, 2014).  It is plotted in units of $L/4\pi D_{min}^2 = L/4\pi (R_E-R)^2$ where the value of $R_E$ for Earth is taken to be 8.5 kpc.  The middle panel displays the sky-averaged flux per steradian in units of $L/4\pi D_{min}^2$, and for $ R=9.5 $ kpc, its value is 0.11.  Consider the example of a Galactocentric ring source at $R=9.5 $ kpc. Then $4\pi D_{min}^2\approx1.2 \cdot 10^{44}$ cm$^2$, and a source luminosity at $E$=2.4 EeV of $dL(E)/d\ln E|_{E=2.4 {\rm EeV}} = 1.2 \cdot 10^{35}$ erg  s$^{-1} \rm sr^{-1}$ implies an energy flux of  $F = 1\cdot 10^{-10}$ erg cm$^{-2}$ s$^{-1}$ sr$^{-1}$, the value implied by the lower panel of figure 1 [see also Eichler and Pohl, (2011)]. The implied total luminosity in UHECR above $10^{17}$ eV is about 3 times this value, or about $10^{35.5}$ erg/s.
The anisotropy,  relative to a radial vector from the Galactic center, of UHECR coming from each ring is plotted in the lower panel.  The red line indicate the observational upper limits (less than 2.3\% in the 2-4 EeV energy band, Pierre Auger Collaboration, 2011) on the absolute value of the anisotropy of the overall UHECR flux, which, in reality, is from a superposition of sources with varying $R$. The black line indicates time and source averaged anisotropy (1.1\%), taking into account a source spatial distribution that matches the star formation rate (Kumar and Eichler, 2014).

A source of uncertainty in estimating $L$ is the choice of $D_{min}$.
The example $R$=9.5 kpc is motivated by the consideration that  rings with radii between 6.5 and 10.5 kpc  are most likely to contribute so the mean values 7.5 kpc and 9.5 kpc should be  reasonably well represented at having contributed to the present flux. The reported anti-center anisotropy at 8 EeV suggests in the context of the present model that the most recent nearest source(s) would have been mostly outside the solar circle. 
It should be clear, however, that the discreteness of the sources and the uncertainty in the their locations makes this estimate fundamentally uncertain, and it is presented merely to show that the energy demands are reasonable. 
Choosing $R= 6.5$ kpc (10.5 kpc) raises the estimate of $L$ by a factor of approximately 8 (2).

The time average prompt emission luminosity of Galactic GRB, if they have the same luminosity per unit mass $\dot w $ as the spiral galaxies in the rest of the universe, $\dot w \sim 6 \cdot 10^{42}$ erg Mpc$^{-3}$ yr$^{-1} /\rho_{sp}$, where $\rho_{sp}\sim 10^{-30} \rm g\, cm^{-3}$ is the mass density in spiral galaxies, would be $M_G \cdot \dot w \sim 1.3\cdot 10^{37}$ erg s$^{-1}$. So the energy needed for UHECR at 1 EeV is     below  the expected energy budget for prompt emission.  (Here we have assumed that the mass in spiral galaxies, including the dark halos, is about $10^{-1}$ of the critical density $\rho_c \sim 1 \cdot 10^{-29}\rm g\, cm^{-3}$, and that $M_G = 10 ^{12} M_{\odot}$.)

The spectrum goes roughly as ${dN/ dE} \propto E^{-3}$, beginning at $E= 10^{17}$ eV until the ankle at $\sim 4 \cdot 10^{18}$ eV. At the low energies,  up to a few $10^{18}$ eV, the spectrum is dominated by escaping neutrons, coming mostly from photodissociation of heavy nuclei from within the fireball\footnote{For high luminosity GRBs, as it is the case considered here, proton acceleration is limited by photomeson production during the early stage of the shock propagation, and contribute to neutron production (although the nuclei contribution is largely dominant).}, and little of the energy in escaping particles goes below that. The composition clearly goes from proton-dominated at low energies to trans-iron dominated at large energy, compatible with the trend suggested by Auger data (Pierre Auger Collaboration, 2015).

\section{Discussion}

We have considered the conventional possibility that UHECRs are extragalactic and have a very hard spectrum because all but the highest energy CR are trapped and adiabatically cool near the source.
The problem with this scenario is that the energetic requirements imposed on the GRB are then two to three orders of magnitude\footnote{Depending on what minimal energy is considered. The cosmic rays in escaping CR alone require only a factor of $\sim 10^2$ times the prompt emission. But there may be a bolometric correction applied if the escaping particles comprise only a fraction of the total cosmic ray output of the shock.}  greater than the prompt emission of GRB (Eichler, Guetta and Pohl, 2010,  Globus et al. 2015), whereas the afterglow in any given GRB gives no evidence of the blast energy being larger than the prompt emission.  

If UHECRs are extragalactic, the ankle-like feature observed in the cosmic-ray spectrum at $\sim 4 \cdot 10^{18}$ eV can be attributed  to the GCR/EGCR transition (e.g. Globus, Allard and Parizot 2015 for a recent account), where the end of the Galactic cosmic-ray component  takes place at the ankle.
In the Galactic scenario proposed here, the ankle appears because of a  softer  (than the other ion species) ex-neutron component in the UHECR spectrum and also the extra $Q^{\alpha}$ factor, corresponding to the rigidity-dependent residence time in our Galaxy. 
  Therefore, no additional Galactic component needs to be postulated below the ankle, down to  $\sim10^{17}$ eV. 

Recently, the  KASCADE-Grande collaboration reported an ankle-like break in the light cosmic-ray spectrum at  $\sim10^{17}$ eV (Apel et al., 2013).
In our scenario, this light ankle is interpreted as a transition between two light Galactic components: a first light Galactic component extending up to  $\sim10^{17}$ eV and a second light Galactic component coming from the output of our Galactic GRB, that is clearly dominated by the protons between  $10^{17}$ eV and  $\sim10^{19}$ eV.
The softer ex-neutron component is a generic feature of nuclei acceleration in photon-rich environments (we considered here GRB internal shocks as acceleration sites). If this is indeed the case for UHECRs sources, both the Galactic and extragalactic models predict a predominantly light component below and across the ankle, in good qualitative agreement with the recent results of KASCADE-Grande and Auger.\footnote{Auger  recently reported that the relative fraction of protons becomes smaller below $10^{18.3}$ eV with $<\ln A>\sim 2$ at $10^{17}$ eV (Apel et al., 2013). It should be noted that in our scenario we need to assume another Galactic component at lower energies, with a cut-off for the protons at  $\sim10^{17}$ eV (to fit the light ankle). If all nuclei have the same rigidity spectrum, we expect that if some Galactic sources accelerate protons up to $\sim10^{17}$ eV, they  accelerate iron up to $2.6\cdot 10^{18}$ eV, though the heavy metal content  of the thermal gas need not be as high as  within the GRB fireball.}
At trans-ankle energies, the evolution of the UHECR composition implied by our source model is also in very good qualitative agreement with the trend suggested by Auger composition analyses (Pierre Auger Collaboration 2015): the composition is dominated first by protons, then by intermediate, and finally by heavy nuclei. Such a trend is a natural consequence of the acceleration process.
In other words, the models differ only in the location of the GRBs, whether the UHECR flux is dominated by extragalactic or by Galactic GRBs. 

A major concern for the Galactic model is the remarkable isotropy displayed  by EeV cosmic rays, as well as the inward (towards the Galactic center) anisotropy displayed at somewhat higher energies (Aartsen et al., 2015),   as it  is often assumed that Galactic of UHECR would give a larger, outward  anisotropy. 

However, this needn't be the case if the Galactic magnetic field has an equatorial current sheet.  Kumar and Eichler (2014) predicted an anticenter flux (i.e. toward the center) at sufficiently high energy when there is a current sheet that the UHECR drift in along.  
Such a  current sheet results from the presence of a current sheet at the magnetic equator associated with the
reversal of the sign of the toroidal field.  UHECR ions execute $\nabla B$ drift at this current sheet inward, while eventually drifting outward along the magnetic poles. A typical drift trajectory is shown in Figure 3. 
A particle distribution after 30 scattering times is shown in Figure 4. It is clear that UHECR of sufficiently high gyroradius injected at a finite Galactic radius on the equator tend to drift in along the current sheet even while diffusing outward down a source density gradient. This can be seen in the indentation of the distribution on the current sheet. 
The implication is that UHECR (of sufficiently high energy) that originate well inside the solar circle are unlikely to reach Earth, because they drift out of the Galactic plane first, while those originating from sources outside the solar circle are more likely to reach Earth.\footnote{This effect has also been recently reported by  Farrar et al.  (2015) at 0.3 EV (Figure 6).} This results in an inward anisotropy at sufficiently high energy. The low observed anisotropy at the ankle could then be because this is near the energy at which the anisotropy reverses sign.

A prediction of this explanation for the low anisotropy is that it should depend on rigidity.  Light and heavy nuclei, as much as they can be distinguished by airshower analysis,  should display different anisotropies at the same energy. 

Finally, we emphasize that, in  setting to be proportional to the UHECR flux to the average production spectrum divided by  the escape rate, we have made a statistical assumption. In fact, the UHECR spectrum, if they originate in GRB, may have a time variable spectrum and a time variable anisotropy at Earth due to intermittency effects (Kumar and Eichler, 2014). Future work should address this point. 

We acknowledge support from the Joan and Robert Arnow Chair of Theoretical Astrophysics, and support from the Israel-U.S. Binational Science Foundation and the Israeli Science Foundation, including an ISF-UGC grant.  NG acknowledges the support of the I-CORE Program of the Planning and Budgeting Committee, the Israel Science Foundation (grant 1829/12) and the Israel Space Agency (grant 3-10417), and Denis Allard for useful conversations.

\begin{figure}
    \centering
    \includegraphics[width=0.6\textwidth]{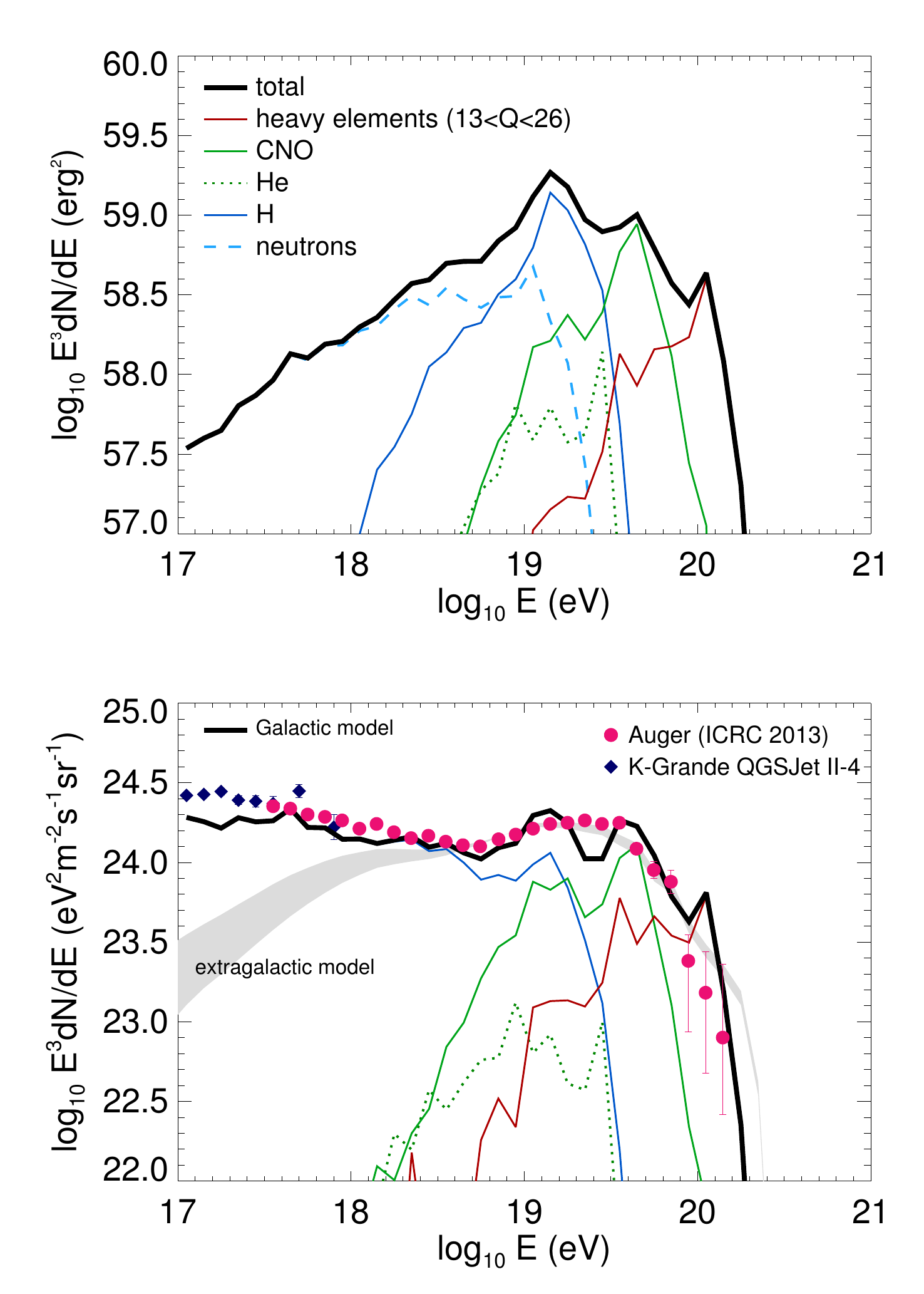}
    \caption{Upper panel: the CR output of a high luminosity GRB (Globus et al. 2015). Lower panel: The expected cosmic ray flux, adjusted to the Auger and KASCADE-Grande data, assuming that cosmic-rays escape from the Galaxy with an energy dependent escape rate that is $\propto (E/Q)^{-\alpha}$ with $\alpha \sim 0.9$. 
The grey line shows the UHECR spectrum assuming that the sources are extragalactic GRBs (the plot is taken from Globus, Allard and Parizot 2015). The spread shows different evolution of the source density as function of redshift $z$: $(1 + z)^{\beta}$, with $2.1< \beta < 3.5$, $\beta=2.1$ is the GRB cosmological evolution predicted by Wanderman and Piran (2010). } 
\label{figure1}
\end{figure}

\begin{figure}
   \centering
    \includegraphics[width=0.9\textwidth]{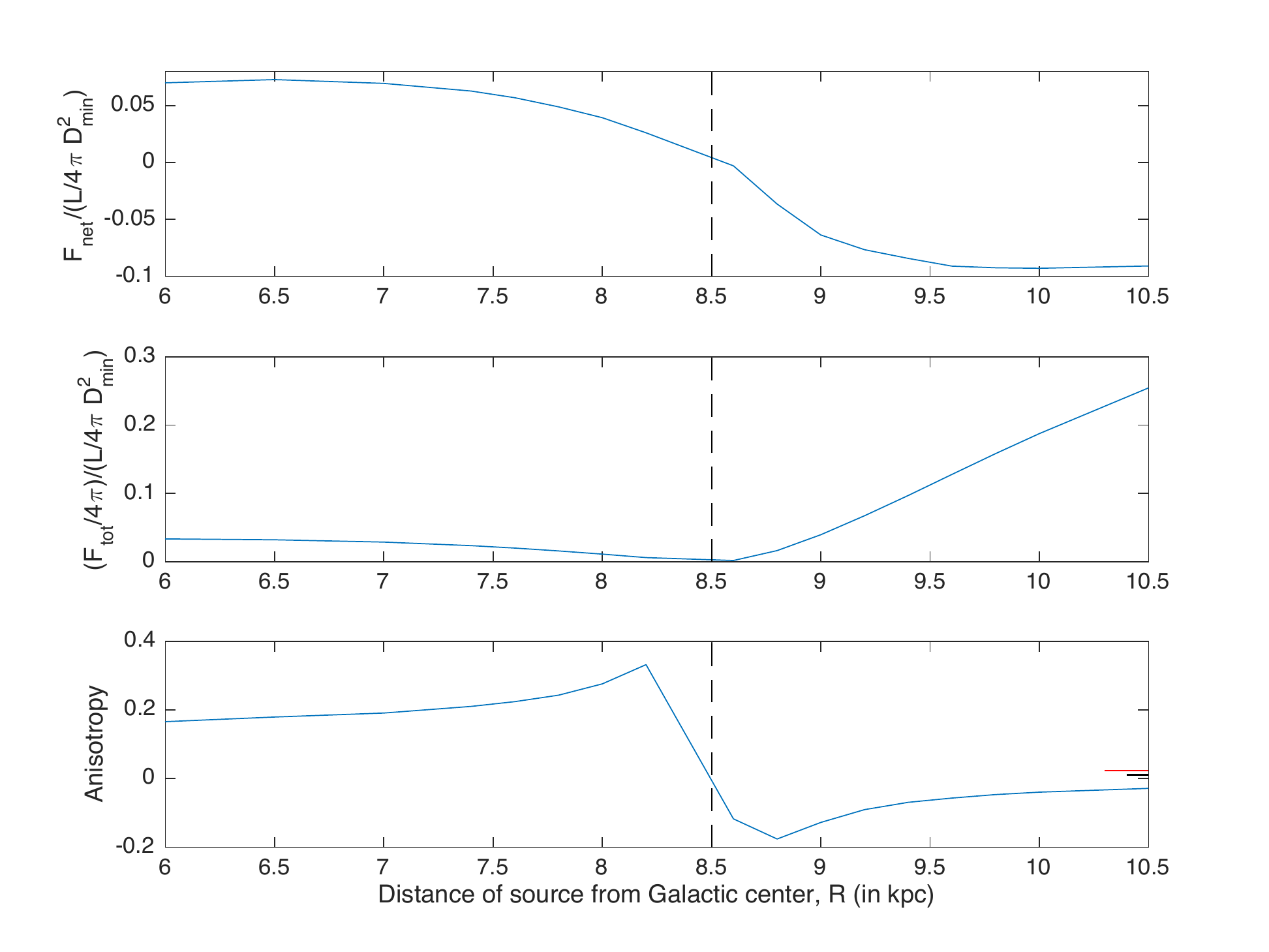}
\caption{Upper panel: the flux at Earth is plotted, given the luminosity $L$ of a ring source, as a function of the Galactocentric radius $R$, in units of the Earth's distance $D_{min} = R_E - R= 8.5 \,{\rm kpc} -R$, from the closest point on the ring. Middle panel: the sky-averaged flux per steradian in units of $L/4\pi D_{min}^2$. Lower panel: the observed anisotropy at the Earth due to steady ring sources, as a function of the ring radius $R$. The result is based on numerical integration of particle orbits in the Galactic magnetic field, assuming Bohm diffusion of UHECRs. The Galactic magnetic field is assumed to be toroidal, to reverse sign at the equator and to decrease exponentially with $z$ (in this case $B_\phi\propto \,10\mu G\, \exp(-|z|/1.5\,{\rm kpc})$.}
 \label{figure 2}
\end{figure}
\begin{figure}
    \centering
    \includegraphics[width=0.6\textwidth]{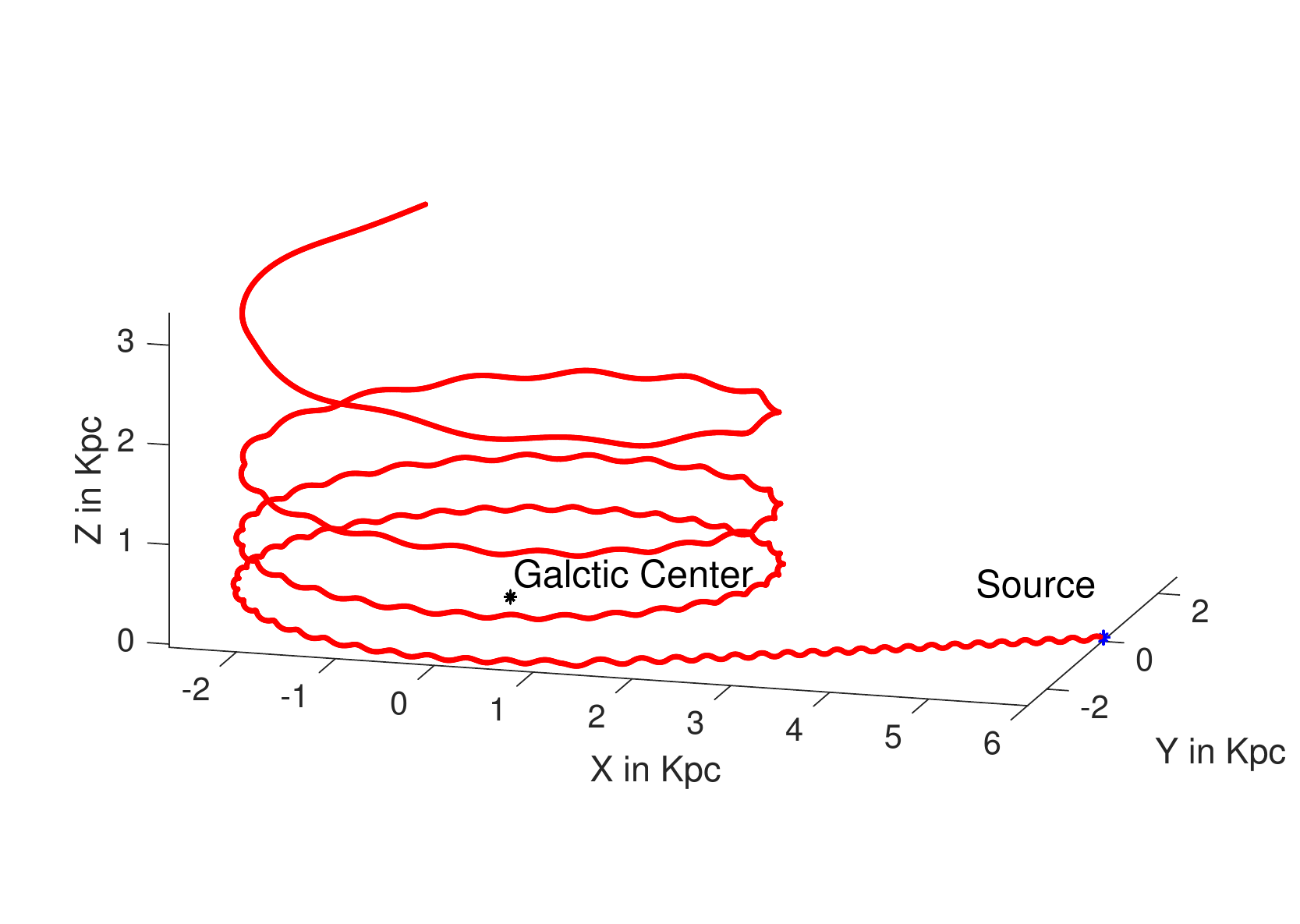}
    \caption{The drift-trajectory of an ultrahigh energy cosmic ray ion with finite angular momentum along the Galactic rotation axis) is displayed in the case of zero scattering. The magnetic field is assumed to completely toroidal and to reverse sign at the equator, meaning that there is a current sheet at the equator.  The UHECR are injected at the current sheet at a radius of 6 kpc. } 
\label{figure2}
\end{figure}

\begin{figure}
    \centering
    \includegraphics[width=0.6\textwidth]{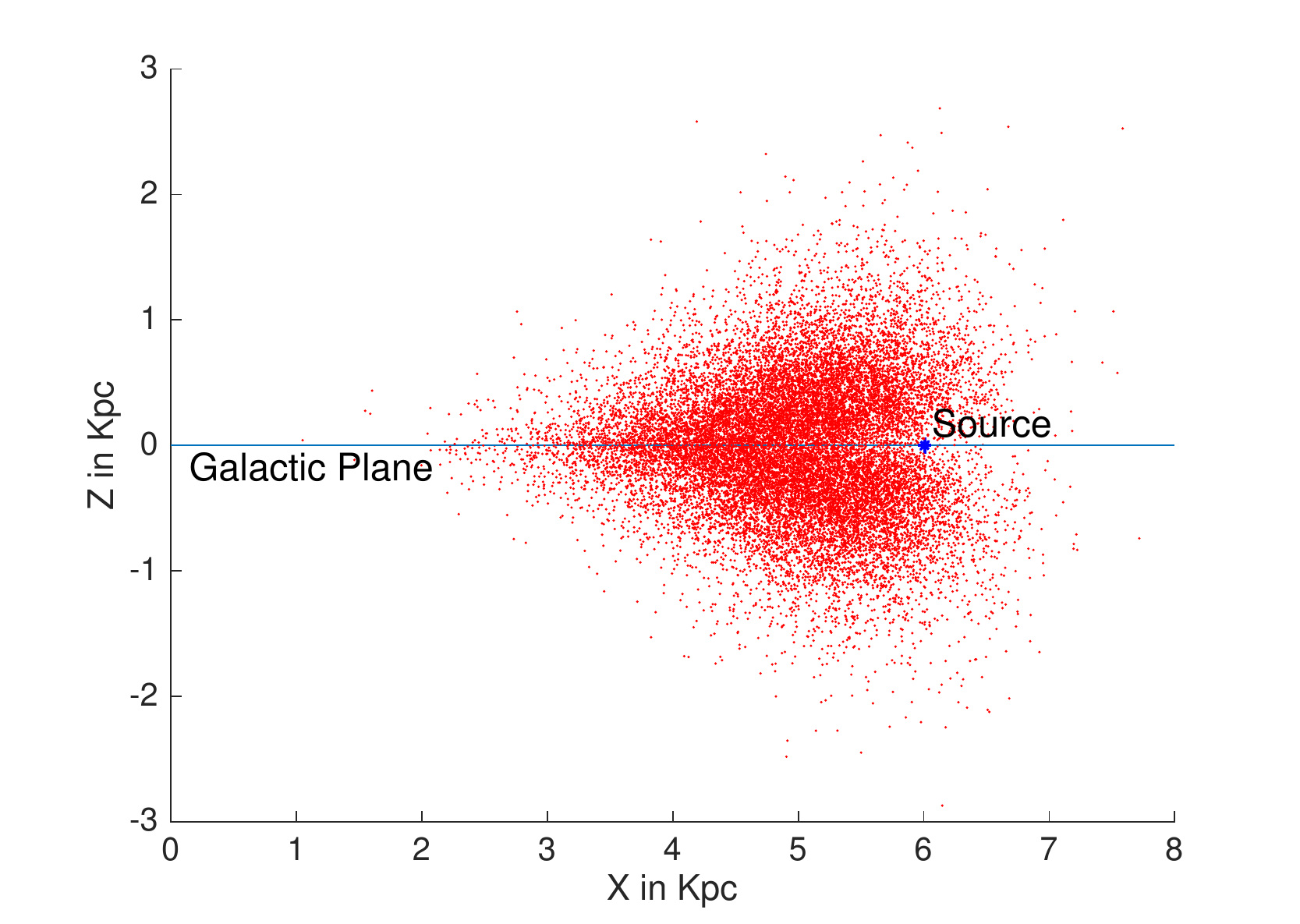}
    \caption{The particle distribution after 30 scattering times with constant injection at the Galactic equator at a Galactic radius of 6 kpc.} 
\label{figure3}
\end{figure}
\end{document}